\documentclass[10pt,conference]{IEEEtran}

\usepackage{amsmath,hyperref}
\usepackage{paralist}
\usepackage{enumitem}
\usepackage{textcomp}
\usepackage{bibentry}
\usepackage{color}
\usepackage{graphicx}
\usepackage{url}
\usepackage{xspace}
\usepackage{multirow}
\usepackage{subfig}
\usepackage{tabularx}
\usepackage{grffile}
\usepackage{subfig}
\usepackage{comment}
\usepackage{booktabs}
\usepackage{balance}
\usepackage{siunitx}
\usepackage[font=small,labelfont=bf,textfont=bf]{caption} 
\usepackage{cite}

\makeatletter
\def\sectionautorefname{\S\@gobble}
\def\subsectionautorefname{\S\@gobble}
\def\subsubsectionautorefname{\S\@gobble}

\makeatother

\begin{document}
\bstctlcite{IEEEexample:BSTcontrol}

\newif\ifdraft

\ifdraft
 \newcommand{\jhanote}[1]{{\textcolor{red}{ ***Shantenu: #1 }}\xspace}
 \newcommand{\iynote}[1]{{\textcolor{brown}{ ***Igor: #1 }}\xspace}
 \newcommand{\mtnote}[1]{{\textcolor{orange}{ ***Matteo: #1 }}\xspace}
 \newcommand{\note}[1]{{\textcolor{blue}{ ***Note: #1 }}\xspace}
 \newcommand{\ian}[1]{{\textcolor{magenta}{ ***Ian: #1 }}\xspace}
 \newcommand{\hm}[1]{{\textcolor{cyan}{ ***Heng: #1 }}\xspace}
 \newcommand{\alex}[1]{{\textcolor{blue}{ ***Alex: #1 }}\xspace}
 \newcommand{\arvind}[1]{{\textcolor{magenta}{ ***Arvind: #1 }}\xspace}
 \newcommand{\hyungro}[1]{{\textcolor{pink}{ ***Hyungro: #1 }}\xspace}
 \newcommand{\fixme}{{\textcolor{red}{ ***Fixme}}\xspace}
\else
 \newcommand{\jhanote}[1]{}
 \newcommand{\iynote}[1]{}
 \newcommand{\mtnote}[1]{}
 \newcommand{\note}[1]{}
 \newcommand{\ian}[1]{}
 \newcommand{\hm}[1]{}
 \newcommand{\alex}[1]{}
 \newcommand{\hyungro}[1]{}
 \newcommand{\arvind}[1]{}
 \newcommand{\fixme}
\fi

\newcommand{\revision}[1]{{\textcolor{blue}{#1}}\xspace}

\title{Coupling streaming AI and HPC ensembles to achieve 100-1000$\times$ faster biomolecular simulations}

\author{
\IEEEauthorblockN{
Alexander Brace\textsuperscript{1,3}, 
Igor Yakushin\textsuperscript{1}, 
Heng Ma\textsuperscript{1}, 
Anda Trifan\textsuperscript{1,4},
Todd Munson\textsuperscript{2}, 
Ian Foster\textsuperscript{1,3}, 
Arvind Ramanathan\textsuperscript{1}}
\IEEEauthorblockA{ \textit{
\textsuperscript{1}Data Science and Learning Division, \textsuperscript{2}Mathematics and Computer Science Division, Argonne National Laboratory, Lemont, USA.}\\
\textit{\textsuperscript{3}Department of Computer Science, University of Chicago, Chicago, USA.}\\
\textit{\textsuperscript{4}Center For Biophysics, University of Illinois Urbana-Champaign, Champaign, USA.}\\
{\normalsize {\{abrace, iyakushin, heng.ma, atrifan, tmunson, foster, ramanathana\}@anl.gov}}}\\

\IEEEauthorblockN{Hyungro Lee\textsuperscript{5}, Matteo Turilli\textsuperscript{5,6}, Shantenu Jha\textsuperscript{5,6}}
\IEEEauthorblockA{ \textit{
\textsuperscript{5}RADICAL-Lab, Rutgers University, New Brunswick, USA.} \\
\textit{\textsuperscript{6}Computational Science Initiative, Brookhaven National Laboratory, Upton, USA.}\\
{\{hyungro.lee, matteo.turilli, shantenu.jha\}@rutgers.edu}}
}

\maketitle

\begin{abstract}
Machine learning (ML)-based steering can improve the performance of ensemble-based simulations by allowing for online selection of more scientifically meaningful computations. We present DeepDriveMD, a framework for ML-driven steering of scientific simulations that we have used to achieve orders-of-magnitude improvements in molecular dynamics (MD) performance via effective coupling of ML and HPC on large parallel computers. We discuss the design of DeepDriveMD and characterize its performance. We demonstrate that DeepDriveMD can achieve between 100--1000$\times$ acceleration for protein folding simulations relative to other methods, as measured by the amount of simulated time performed, while covering the same conformational landscape as quantified by the states sampled during a simulation. Experiments are performed on leadership-class platforms on up to 1020 nodes. The results establish DeepDriveMD as a high-performance framework for ML-driven HPC simulation scenarios, that supports diverse MD simulation and ML back-ends, and which enables new scientific insights by improving the length and time scales accessible with current computing capacity.
\end{abstract}

\section{Introduction}
\label{sec:intro}
The use of molecular dynamics (MD) simulations to explore complex biophysical phenomena such as protein folding and protein-ligand/small molecule docking has transformed understanding of fundamental biology and advanced important applications such as drug design~\cite{Dror_2012}. MD simulations evolve the state of a molecular system by repeatedly computing and updating the position and other properties of individual atoms (or, in so-called coarse-graining methods, sets of atoms) in response to forces imposed by other atoms.  However, conventional MD methods have, for many problems, reached their limits, for two reasons: first, the speed at which even a massively parallel MD code can evolve the state of a particular molecular configuration (measured, for example, in ns/day) is ultimately limited by communication costs (i.e., weak scaling constraints), and second, a simulation may have to overcome local minima to allow the biological system to sample ``interesting events'' which typically occur at much longer, ms to s, timescales~\cite{Dror_2012,Kasson_2018}. 

Several approaches are used to overcome the sampling limitations of conventional MD simulations. A brute-force method would be to design a single-purpose, highly efficient custom supercomputer (e.g., Anton~\cite{Lindorff-Larsen517}) that can access long timescales in a single run. A second approach to overcoming the limitations of conventional MD is to run multiple simulations concurrently, each starting from a different initial state, i.e., as \emph{ensembles}~\cite{balasubramanian2020adaptive,Kasson_2018}. A combination of higher parallel efficiency for individual simulations and the broader exploration of conformational space due to multiple starting points can make such ensemble methods more effective than conventional MD~\cite{zwier_westpa_2015,zuckerman_weighted_2017}. However, the problem of unproductive simulations that fruitlessly explore uninteresting or already visited parts of conformational space remains~\cite{Kasson_2018}.

To understand how we might overcome this latter difficulty, consider how a human expert might approach the problem if they could monitor the progress of individual simulations. Observing that one simulation is exploring an interesting trajectory through conformational space, while a second is stuck in an uninteresting local minimum and a third is mirroring the trajectory of the first, the expert might decide to restart the latter two simulations with different initial conditions. Thus, there is an inherent trade-off between exploration (i.e., identifying new and biophysically relevant/ interesting conformational states explored by the MD simulations) versus exploitation (i.e., leveraging discovered conformational states for accelerating the simulation of biophysical phenomenon of interest, such as protein folding). 

Recent work on machine learning and artificial intelligence (ML/AI)-driven MD ensembles aims, in effect, to use ML/AI methods to emulate the behavior of the human expert~\cite{Smith_2020,wang2020machine,Ribeiro_2018,Ribeiro_2019,Shamsi_2018,Perez_2020}. We see a broad spectrum of approaches in which, for example, one or more ML models are trained once and then used to guide simulations, or alternatively are retrained over time to incorporate new knowledge of simulation progress. Our concern here, however, is not the merits of specific ML approaches, but rather with the methods used to organize ML-guided ensemble computations on large parallel computers and, in particular, with approaches to link simulation, ML training, and ML inference tasks so as to maximize both parallel efficiency and the timeliness of ML guidance, and ultimately the scientific value gained from an ensemble run. 

\emph{Contributions}\/: To this end, we present DeepDriveMD, a framework that allows for the flexible specification of, and scalable execution of, a range of ML/AI-driven MD ensemble simulation strategies.
Distinctive features include:
\begin{itemize}
    \item 
\emph{Flexibility}: DeepDriveMD supports the flexible construction of  ML/AI-driven simulation systems that link popular simulation engines (OpenMM~\cite{openmm_eastman2017openmm}, NAMD~\cite{Phillips2005}, and AMBER~\cite{Case}), ML training, and inference stages with diverse back ends, including TensorFlow and PyTorch (for deep learning), and scikit-learn (other ML tasks).
    \item 
	    \emph{Scalability}: DeepDriveMD enables scaling of these systems to large numbers of ensemble members (both MD and ML tasks) and to extremely large computers and emerging heterogeneous systems. This is implemented by leveraging RADICAL-Cybertools~\cite{balasubramanian2019radical}, which provides a scalable framework to couple diverse simulation and ML/AI back ends within modular workflows, while abstracting the complexity to the application at hand. 
    \item 
    \emph{Online coupling}: DeepDriveMD supports  concurrent execution of the different stages, with high-speed streaming between stages leveraging the ADIOS platform~\cite{godoy2020adios}, and permits rapid and iterative feedback between simulation and ML/AI models (entirely novel in this work).  
\end{itemize}

\begin{figure}[!htbp]
  \centering
  \includegraphics[width=\columnwidth]{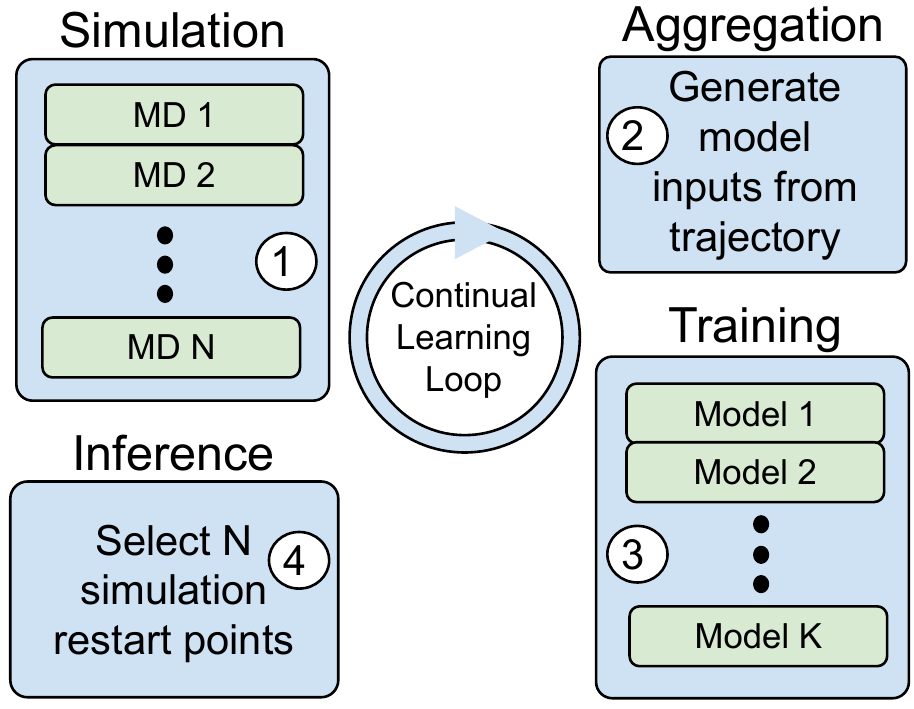}
  \caption{The computational motif implemented by DeepDriveMD to support ML/AI-coupled simulations comprises four stages.
\textit{Simulation}: Simulations are used to explore possible trajectories of a protein or other biomolecular system; 
\textit{Aggregation}: Simulation results are preprocessed for training. 
\textit{Training}: Aggregated trajectories are used to train one or more ML models.
\textit{Inference}: Trained ML models are used to identify conformations for subsequent iterations of simulations. 
  }
  \label{fig:arch_io}
\end{figure}

DeepDriveMD supports, in particular, the computational motif depicted in \autoref{fig:arch_io}. In this motif, MD \emph{simulation} ensembles are run from which distinct data `view's are aggregated. This \emph{aggregation} step pre-processes the simulation data, such as filtering only a subset of atoms of interest, or calculating physical parameters (e.g., root-mean squared deviations (RMSD) to a protein's native state, or more generally a reaction coordinate/collective variable), or simply aggregating conformations from the simulations as they are running. ML/AI techniques are then run across the aggregated data (\emph{training}). Once the ML/AI models are trained, they may be run in \emph{inference} mode to decide which simulations to run next, and/or to terminate less productive MD simulations. (It is also possible to substitute simulations with generative models~\cite{Noe_2020}, or with surrogate models that provide access to new conformations from which simulations can be started.) The resulting \emph{continual learning loop} drives successive iterations of DeepDriveMD simulations.

We use three scientific use cases to demonstrate various DeepDriveMD capabilities and to quantify the benefits that result from its scalability and online coupling. The first use case involves a protein folding simulation, namely that of fast-folding the $\beta\beta\alpha$ (BBA) canonical fold engineered protein, FSD-EY~\cite{Sarisky_2001}, and demonstrates how DeepDriveMD provides 100$\times$-1000$\times$ acceleration, measured in simulation end points/time-to-solution, in sampling low RMSD states that correspond to its fully folded state. We also show that DeepDriveMD can sample 80\% of the BBA folding landscape four orders of magnitude more efficiently than conventional MD simulations. The second use case examines the diversity in MD simulations, typically captured within protein-ligand complexes (PLC). Here, we demonstrate that DeepDriveMD overheads on a large number of compute nodes are minimal, and that the system can thus scale to approximately one quarter of the Summit supercomputer, 1020 of the 4608 full nodes. The third use case demonstrates that DeepDriveMD can scale to large ($O(10^7-10^8)$ atoms) systems and can be more effective than conventional MD in capturing `rare' events, such as spike protein attaching to host receptor binding domain~\cite{casalino2020aidriven}. Taken together, these three use cases demonstrate how DeepDriveMD provides a flexible and scalable framework for supporting ML/AI-coupled ensemble MD simulations while highlighting the performance tradeoffs involved in supporting these use cases.

\section{Related Work}
\label{sec:related}
Ensemble MD sampling methods have recently been augmented with ML/AI techniques~\cite{Ribeiro_2018,rydzewski2020multiscale,Bonati17641,Hernandez_2018,jung2019artificial}, including reinforcement learning and other complementary approaches~\cite{RAMANATHAN2021216}. However, these previous studies have focused primarily on prototypical systems such as small peptides/proteins to demonstrate this general workflow's feasibility.

Previous tools for building ML-coupled simulation workflows have focused on integrating popular ML backends, TensorFlow and PyTorch, with simulation toolkits~\cite{korshunova2021openchem,Doerr_2021,Ramsundar-et-al-2019}. These frameworks mainly provide a configurable API to specify workflow parameters that optimize the use of parallel computing resources. Parallel scripting systems have also been used~\cite{wilde2009parallel}.  Tools such as SmartSim~\cite{smartsim} incorporate an HPC job scheduler to orchestrate task assignment and resource allocation through Slurm, Cobalt, or PBSPro. Within multiscale biomolecular simulations, the recently developed MuMMI~\cite{di2019massively} framework uses ML techniques to link  multi-scale models while simultaneously optimizing HPC resource utilization. Other frameworks, such as NVIDIA SimNet~\cite{hennigh2020nvidia} enable AI-driven acceleration of forward and inverse problems in multi-physics (e.g., turbulence), with a particular focus on solving partial differential equations. Tools such as Proxima~\cite{zamora2021proxima} and Colmena~\cite{Colmena2021} also represent advances in managing large AI-enabled HPC workflows.  Many recent developments focus on accelerating drug  discovery workflows~\cite{samadejacobs2021high,glaser2021high,saadi2020impeccable,GORGULLA2021102021}. There are three distinct computationally expensive phases: (1) generating the simulation data, (2) training the model, and (3) using the model to guide simulations to generate scientifically meaningful data. The cost of the first dominates in the use cases that we consider in this paper, but in other cases, powerful ML training tools such as KubeFlow~\cite{bisong2019kubeflow}, MLFlow~\cite{zaharia2018accelerating}, and LBANN~\cite{essen2015lbann} are available.

\section{Our Science Drivers}
\label{sec:sdrivers}
We provide more details on the science use cases introduced earlier. Each involves a real scientific problem that the DeepDriveMD team and its collaborators have studied while developing methodology and infrastructure; each also permits evaluation of a different aspect of DeepDriveMD performance.

\subsubsection*{\textbf{UC1: Protein folding}}\label{ssec:protein_folding}
Protein folding refers to the process by which a protein chain is translated to its native three-dimensional structure. Despite recent advances in purely ML-based structure prediction~\cite{jumper2021highly}, computationally demanding simulation methods remain important to characterize the partially folded or misfolded intermediate states (so that these states can be targeted via small molecules for drug discovery). To evaluate to what extent our ML techniques can accelerate simulations by enabling more effective sampling of folding events, we apply DeepDriveMD to the problem of folding FSD-EY (PDBid: 1FME), a small (505 atom, 28 amino-acid residue) protein that adopts a canonical $\beta\beta\alpha$ (BBA) fold. We choose FSD-EY because we can compare the scientific performance of our ML-integrated workflow, measured in terms of the fraction of sampled states visited per unit time, against long-timescale simulations with Anton-1~\cite{Lindorff-Larsen517}
hardware. 

\subsubsection*{\textbf{UC2: Protein Ligand Complex}}\label{ssec:plc} 
Protein ligand complex (PLC) computations are used to evaluate whether and how a small molecule (a ligand) may bind with a protein. In this use case, we perform PLC simulations for multiple ligands against the papain-like protease (PLPro) binding site of the SARS-CoV-2 virus. The solvated PLPro system has $\sim$132K atoms and 309 residues. These simulations are performed within a single DeepDriveMD run, with all simulation results aggregated to train a single ML model as a shared representation of the entire ligand search space. (This application is part of an ongoing computational campaign that aims to discover novel molecules that can inhibit SARS-CoV-2. If a specific ligand stabilizes the protein when bound, then scientists can conclude that it may interact with, and potentially inhibit, PLPro.) 

\subsubsection*{\textbf{UC3: Many-Atom Multiscale System}}\label{ssec:mas} 
The need to study increasingly complex biological systems has also motivated the development of multi-scale simulations, where information from one simulation scale (e.g., atomistic) is transferred to a different scale (e.g., coarse-grained), or between different spatial scales (e.g., individual spike protein vs.\ spike proteins embedded within a whole virus). These simulations also use ML methods to transfer information across scales; however, given the many simulations to be carried out, and the large expense of simulating these $O(10^7-10^8)$-atom biomolecular systems, we did not use this workflow as a formal use-case but proffer it as an illustrative example~\cite{casalino2020aidriven}. For UC3, we present a vignette of performance analysis from this illustrative example~\cite{casalino2020aidriven}, where we examine the performance trade-offs involved in simulating a fairly large biological system, consisting of the SARS-CoV-2 Spike protein (approximately 2 million atoms). 

\section{DeepDriveMD Design and Implementation}
\label{sec:design}
We now describe how the ML-coupled-with-HPC simulation motif of \autoref{fig:arch_io} is realized within DeepDriveMD, describing first the overall design and then the implementation.

\subsection{Design}\label{ssec:design}

DeepDriveMD is intended to serve as a general and extensible framework for the scalable, high-performance ML-guided simulation of proteins and other biomolecular systems. Any DeepDriveMD application, including those constructed to address our use cases UC1, UC2, and UC3, combines MD simulations, ML training, and ML inference components  (the Simulation, Training, and Inference stages in \autoref{fig:arch_io});  it may also include an optional Aggregation stage. Typically, these stages execute repeatedly, with results from each new set of simulations used to (re)train ML models that are then used to establish the next simulations. As we discuss in the following subsection, the stages can be run sequentially (one after the other), but in general we want them to run concurrently so as to maximize concurrency and timeliness of the information used for ML training.

DeepDriveMD can thus support a wide variety of applications, each of which may employ a different MD code, ML model, and ML training method (each with potentially widely varying computational and data requirements) and run at different scales (e.g., from a handful to thousands of nodes on an HPC system). It must deal with a large dynamic range in the (1) number and scale of the simulations to be run, including the substitution of simulation codes by less expensive ML surrogates, and (2) frequency and degree of coupling between simulations and ML models. DeepDriveMD can also operate on a variety of different HPC platforms and integrate MD and ML codes implemented with different technologies, including OpenMP and MPI, while also allowing for scaling via large-scale task parallelism, for example when running multiple concurrent MD, ML training, and ML inference tasks.

To support this diversity of requirements, DeepDriveMD is implemented as an extensible software system that supports the generalized coupling of ML with HPC simulation tasks for a range of frequencies and volumes. Thus the user can specify, for example, the number of ligands; the number and type of MD simulations; the number and type of ML models to train on simulation outputs; and the methods used to feed data to ML models, to make inferences via one or more of those models, and to use inference outputs to drive MD simulations.

\begin{figure}[!htpb]
  \centering
  \includegraphics[width=0.5\textwidth]{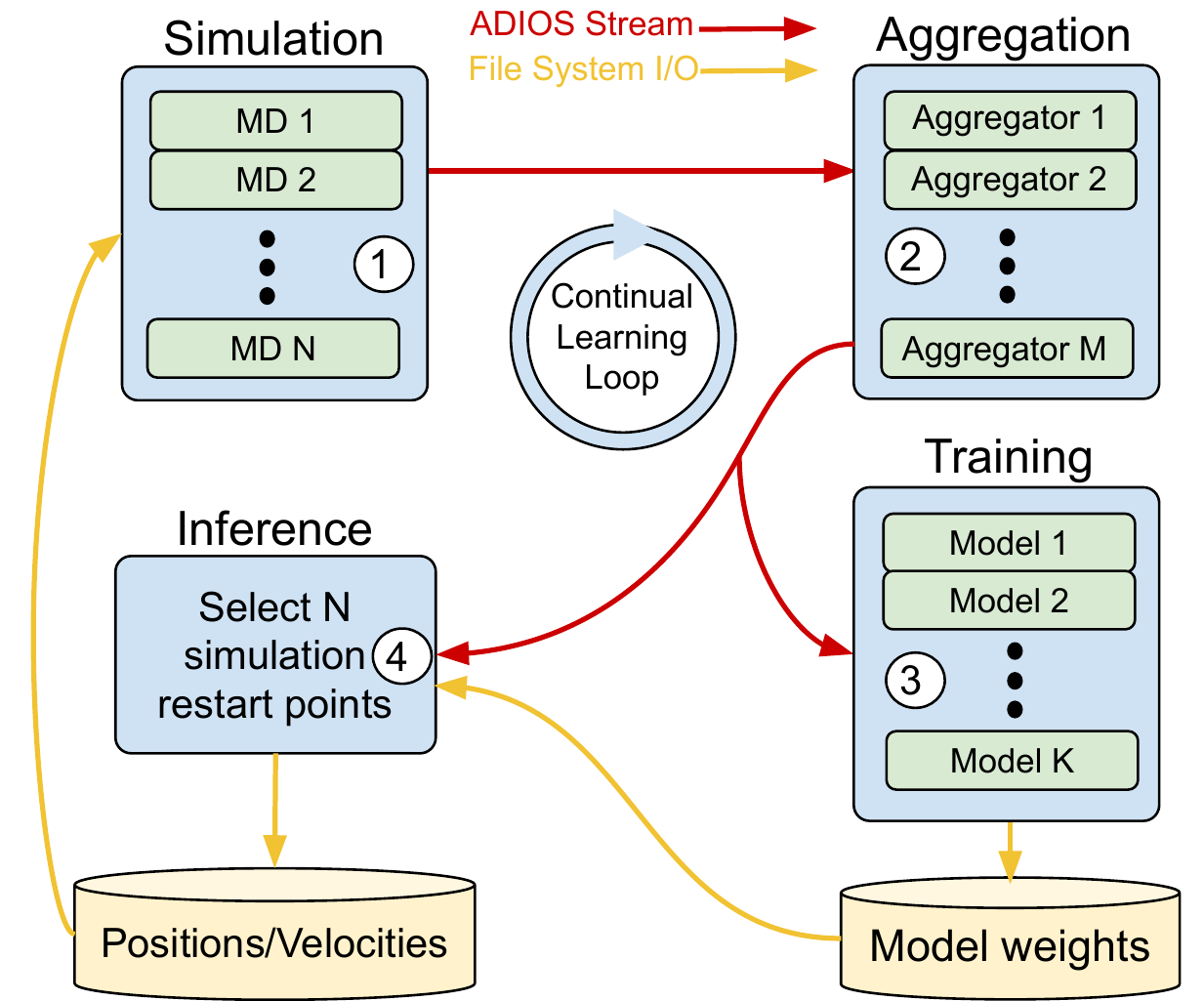}
  \caption{DeepDriveMD architecture.
  In blue are DeepDriveMD components; green
  are tasks, managed by RCT; red are ADIOS streams; yellow is the file system.
  All tasks run concurrently.}
  \label{fig:streaming_DDMD3}
\end{figure}

\subsection{DeepDriveMD Implementation}
\label{sec:implementation}

The DeepDriveMD implementation uses the RADICAL-Cybertools (RCT) Python packages to initiate and manage the execution of the various tasks that comprise the different components. RCT components~\cite{balasubramanian2018harnessing} provide powerful methods and tools that abstract many details associated with mapping many-task applications onto heterogeneous computing systems, for example by enabling interoperability across heterogeneous distributed infrastructures, and implementing pilot job mechanisms that allow users to submit pilot jobs to computing infrastructures and then use the resources acquired by the pilot to execute workloads~\cite{turilli2018comprehensive}. 

Specifically, DeepDriveMD uses the RADICAL-Ensemble Toolkit (EnTK), designed to support the programming of applications comprised of ensembles of tasks. Tasks may be grouped into stages and stages into pipelines, and EnTK executes tasks concurrently or sequentially, depending on their priority relations. Tasks managed by EnTK may be algorithmically heterogeneous (e.g., simulation, aggregation, training, inference); use different parallel computing methods (e.g., threads, OpenMP, MPI, process pools); require different types and amounts of resources (e.g., CPU cores, GPUs, amount of RAM or file system space); and take different amounts of time to execute.

Using the EnTK programming model, we have developed two implementations of DeepDriveMD: (1) DeepDriveMD-F used for performance characterization, and (2) DeepDriveMD-S, an optimized version for which we report scientific validation of UC1 and UC2 as well as performance analysis. DeepDriveMD-F executes the four stages shown in \autoref{fig:arch_io} in sequence, with each completing before the next begins; DeepDriveMD-S runs them continuously and asynchronously, constrained only by data flow synchronization. Thus: (a) the Simulation component performs simulations and streams results to the Aggregation component (as shown in \autoref{fig:streaming_DDMD3}, these two components comprise $N$ and $M$ EnTK Tasks, respectively; invariably, $N>M$ and each simulation task is linked to one Aggregation task); (b) each Aggregation task collects preprocessed data from its subset of simulation tasks and streams it to the Training component; (c) the Training component receives any pending data from the Aggregation component, (re)trains its ML model(s) in an online fashion with the latest data, and communicates updated model weights to the Inference component; and (d) the Inference component uses data from the Aggregation and Training components to make decisions as to whether to continue or terminate a running simulation from the ensemble, and assign the initial configurations for new simulation tasks in the Simulation component. Each component thus has its own independent runtime loop, and is terminated only when runtime is exhausted.

To facilitate communication between components, the DeepDriveMD-S implementation uses the Adaptable Input Output System (ADIOS)~\cite{godoy2020adios}. ADIOS enables interprocess communication via file or network without changing the application code. The native storage format used by ADIOS is a binary-packed (BP) file, similar to HDF5 ~\cite{folk2011overview}. When communicating via network, the Sustainable Staging Transport (SST) protocol supports blocking and non-blocking communication. In this work, we used blocking communication with a fixed size buffer to force producers to wait for consumers to read data before writing more. In UC1 and UC2, we used network communication between simulation and aggregation tasks and file communication from aggregation tasks to training and inference tasks. However, to optimize performance in UC3, we switched to network communication between all components. To backup simulation data for post analysis, we use BP files.

\subsection{MD Implementations}
\label{sec:MDML}

We use OpenMM~\cite{openmm_eastman2017openmm}, running on GPUs, to perform all MD simulations described in this paper. 

The BBA protein of UC1 is set up with the Amberff99SB-ILDN force field~\cite{amber99_lindorff2010improved}, using the implicit GBSA-OBC solvation model~\cite{implicit_sol_onufriev2004exploring}, while the PLpro protein of UC2 is set up with the Amberff14SB force field~\cite{amber14_maier2015ff14sb}, using the TIP3p water model. Other details of the simulation set up are provided in prior publications~\cite{Lee_2019,saadi2020impeccable}. 

The simulation set up for UC2 was identical to \cite{Bhati_2021}. A total of 120 top-ranking ligands from a high-throughput virtual screening campaign~\cite{Clyde_2021} was used to start of the simulation campaign. However, as the simulations progress, the ML method (see below) can automatically pick up ``interesting'' conformations that correspond to increasing the protein-ligand heavy-atom contacts. The increase in the number of contacts usually implies that the ligand is stabilized within a binding pocket of the protein, which is a property of interest for designing small molecule inhibitors.

\subsection{ML Implementations}

We use an unsupervised ML model, namely, the convolutional variational autoencoder (CVAE)~\cite{Bhowmik_2018} written in Keras/TensorFlow 2.1.2 \cite{chollet2015keras, tensorflow2015-whitepaper}. The CVAE automatically reduces the high dimensional MD trajectory data into a latent vector representation in which similar energetic and structural states cluster together.

The CVAE views the MD trajectory as a contact matrix of $C^\alpha$ atoms (within 8 \AA{} cut-off) at each frame, and learns to represent this matrix in a latent space~\cite{Bhowmik_2018}. The model architecture consists of a symmetric encoder/decoder pair with four convolutional layers. We adjust the convolutional layers based on the size of the system. For smaller systems (e.g., contact maps of size 28$\times$28) we use 64 filters with a kernel size of three for all layers, and a stride of two in the second layer whereas for large systems (e.g., 309$\times$309 contact maps), we use strided convolutions in all layers and 32 filters with a kernel size of five in the first layer. We then follow the convolutional layers with a single linear layer of 128 neurons and dropout of 0.4. The latent space is fixed at 10 dimensions and the decoder, composed of transposed convolution operators, reconstructs the input contact matrix. We define the loss function as the sum of the binary cross entropy reconstruction and KL divergence to an isotropic Gaussian prior $\mathcal{N}(0,1)$. This loss function is optimized using RMSprop with learning rate 0.001, $\rho= 0.9$, and $\epsilon= 1e-08$.

\subsection{Inference/ Outlier detection}
The purpose of the Inference component is primarily to select interesting conformations to restart new simulations from. Traditionally, this selection constitutes a biophysical quantity of interest, i.e. a reaction coordinate, that can be tracked as the simulations run~\cite{REAP_shamsi2018}. However, such reaction coordinates are generally system specific and not always known \emph{a priori}. To circumvent this issue, we opt to use the CVAE approach to embed the conformations in a low dimensional manifold. As we have demonstrated before, these embeddings can capture biophysically relevant reaction coordinates~\cite{Bhowmik_2018}. 

In order to search for undersampled regions of the conformational space, we use
traditional outlier detection methods on the latent embeddings produced by the CVAE. Since the CVAE tends to form well-defined clusters, we found
it best to use Density-Based Spatial Clustering of Applications with Noise
(DBSCAN)~\cite{Ester96adensity-based}, as implemented in
RAPIDS~\cite{RAPIDS}, to discover outlier points away from the main clusters. These points represent potentially rare conformations which, if sampled with more MD simulation, could advance the search space. To further refine these outliers, we then employ any known biophysical reaction coordinates to optimize the selection, thus combining the data-driven choice with expert knowledge.

For UC1, we chose the root mean squared deviation (RMSD) to the native state of the protein as a reaction coordinate to track the progress of our simulations and filter the outlier selection. The native state represents a structure, determined experimentally via either X-ray crystallography or nuclear magnetic resonance (NMR), against which simulation progress is usually measured. For the BBA-fold, the NMR ensemble has an average RMSD of 1.3 $\pm$ 0.5~\AA, representing a narrow definition of the native state for this protein. 

For UC2, we expected that at the end of the DeepDriveMD run, the outlier selection would automatically `learn' to keep only stable ligands within the binding pocket of the protein while getting rid of others that are not as stable. To filter the set of DBSCAN outliers,  we use the Local Outlier Factor (LOF)~\cite{LOF} algorithm, as implemented in scikit-learn~\cite{scikit-learn}, to pick the most distant outliers returned from DBSCAN.

\section{Scientific Validation}\label{sec:sci_app}
In light of the methodological novelty of ML-driven ensemble simulations, we are concerned to assess the biophysical (scientific) validity of our results.
\begin{figure*}[ht]
    \centering
    \subfloat[ML, no RMSD considered]{
        \includegraphics[height=3.0cm,trim=2mm 0 2mm 0,clip]{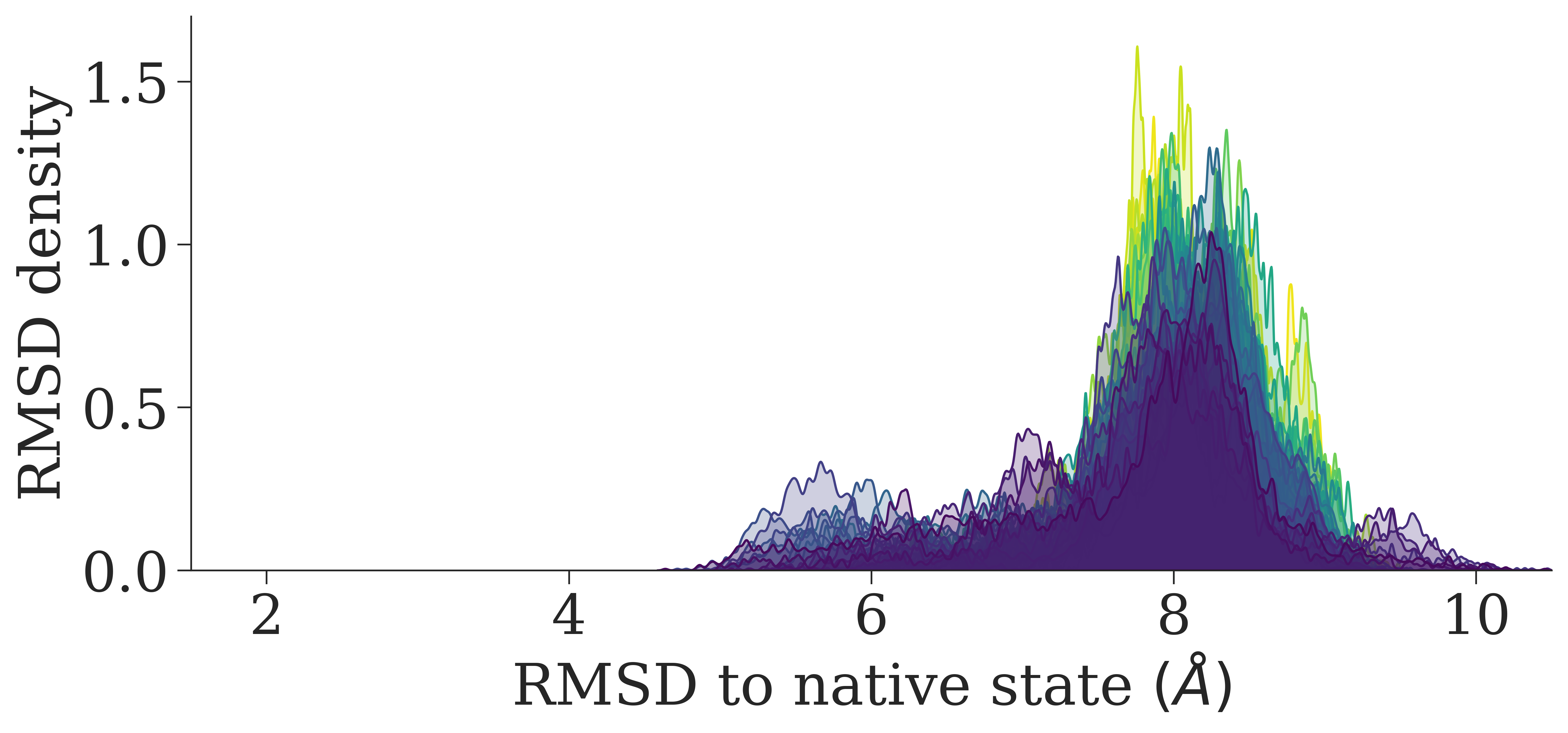}
    }
    \subfloat[no ML, greedy RMSD]{
        \includegraphics[height=3.0cm,trim=28mm 0 2mm 0,clip]{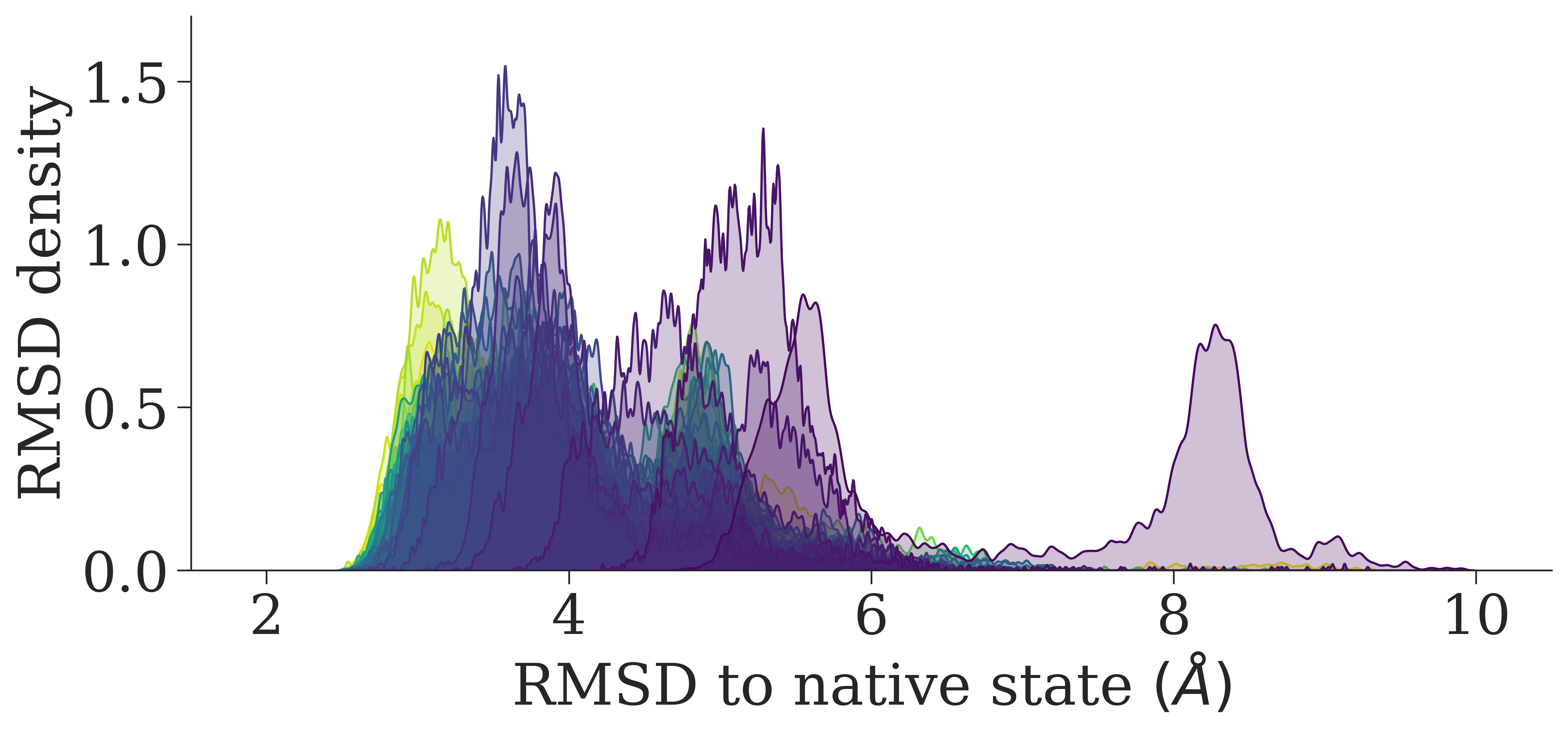}
    }
    \subfloat[ML + RMSD]{
        \includegraphics[height=3.0cm,trim=28mm 0 2mm 0,clip]{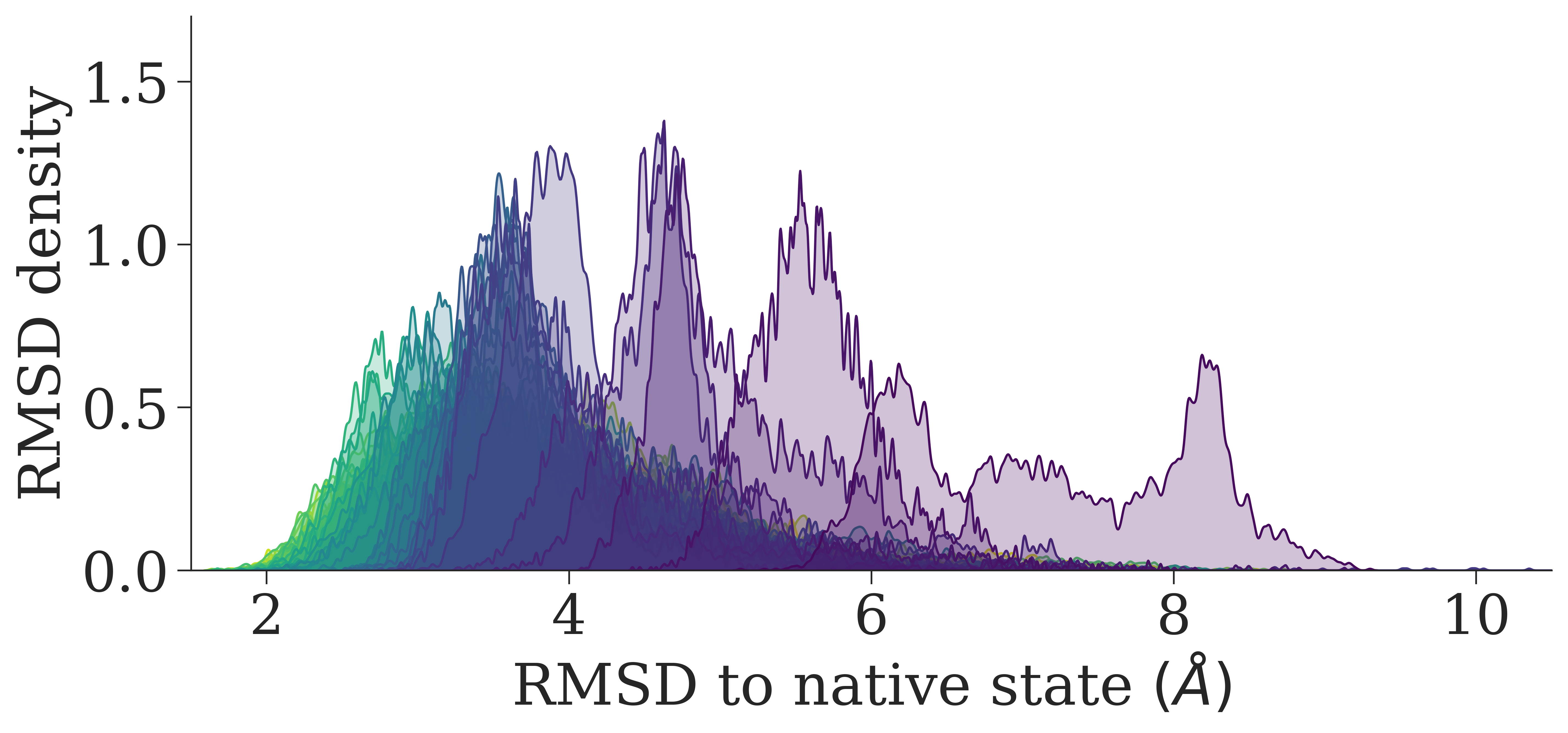} 
    }
    \caption{Comparing BBA folding pathways via the RMSD (\AA) to native state evolution over each iteration of DeepDriveMD, progressing from purple/early iterations to yellow/late iterations, when using three different outlier detection methods. For (a), we see no distribution shift towards smaller RMSD values, as would be indicative of BBA folding; ML without RMSD is not effective. 
    In (b) and (c), we see a clear distribution shift, with (c) performing the best, ultimately achieving a minimum RMSD of 1.55 \AA.}
    \label{fig:uc1_science_evolution}
\end{figure*}

\subsection{Validation of UC1}

For UC1, we measure the convergence to the folded state, or how well DeepDriveMD-S simulations 
access the final folded state(s) of the protein as defined in \autoref{sec:implementation}. As the ML progressively learns a latent representation (based on the CVAE) and guides the selection of conformations to be simulated next through successive iterations, we posit that the RMSD of the conformations selected by Inference must progress towards lower RMSDs.  

However, an important question is to examine whether DeepDriveMD-S simulations benefited by the ML approach at all. We therefore performed 10 trials of each of the following experiments (120 parallel simulations for 12 hours duration): 
\begin{enumerate}[label=(\alph*)]
      \item \emph{ML, no RMSD considered}\/: Here, a purely data-driven strategy is employed with no regard to the biophysical reaction coordinate.  Conformations for the next round of simulations are selected based on ML alone, meaning that outliers, found by DBSCAN, are selected without considering the RMSD to the native state.
        \item \emph{No ML, greedy selection by RSMD}\/: Here, only the biophysical reaction coordinate is used and no aspect of the data-driven steering is employed. In this experiment, conformations for the next round of simulations are selected greedily, based only on RMSD. No outliers are used and only the 120 conformations with the smallest RMSD among the last 20,000 conformations are selected.
    \item \emph{ML + RMSD}\/: This experiment uses both the progress coordinate and the ML-based approach to characterize the sampling process. Conformations for the next round of simulations are selected on the basis of both outlier search (with DBSCAN) and then a greedy selection of the best outliers by RMSD.
\end{enumerate}

Since there is some randomness in the pipeline, the best observed RMSD varies between trials. Thus, we ran methods (b) and (c) for 12 hours, 10 times each, with results summarized in Table~\ref{tab:min_rmsd_with_without_ML}. Both the mean and minimum RMSD achieved by (c) are significantly lower than those for (b), leading us to conclude that the ML approach followed by a greedy selection of outliers based on the RMSD reaction coordinate does help in comparison with the pure greedy selection by RMSD.

\begin{table}[!ht]
    \caption{Best RMSD (\AA) in 10 trials for strategies (b) no ML, greedy RMSD and (c) ML + RMSD.}
    \label{tab:min_rmsd_with_without_ML}
    \centering
    \small
    \begin{tabular}{|l|c|c|}
    \hline
           &  \textbf{(b) no ML, greedy RMSD} & \textbf{(c) ML + RMSD} \\
    \hline
      mean   &  2.37$\pm$0.30 & 1.81$\pm$0.21  \\
      min & 1.85 &  1.55\\
      max &  2.93 & 2.20\\
      \hline
    \end{tabular}
\end{table}

We next examined the histograms of the RMSDs from the Inference stage of the simulations, with a color scheme corresponding to the successive iterations of our continual learning loop, with purple representing earlier iterations and subsequent iterations progressing towards green and yellow colors (\autoref{fig:uc1_science_evolution}). As we observe from the plot, ML + RMSD simulations achieve the lowest RMSD, while the others sample subtantially larger RMSD values, indicating the effectiveness of using the learned latent space representation and a biophysically relevant reaction coordinate in driving the sampling of more productive trajectories. 

\subsection{Validation of UC2}

\begin{figure}[!ht]
  \centering
  \includegraphics[width=\columnwidth,height=0.6\columnwidth,]{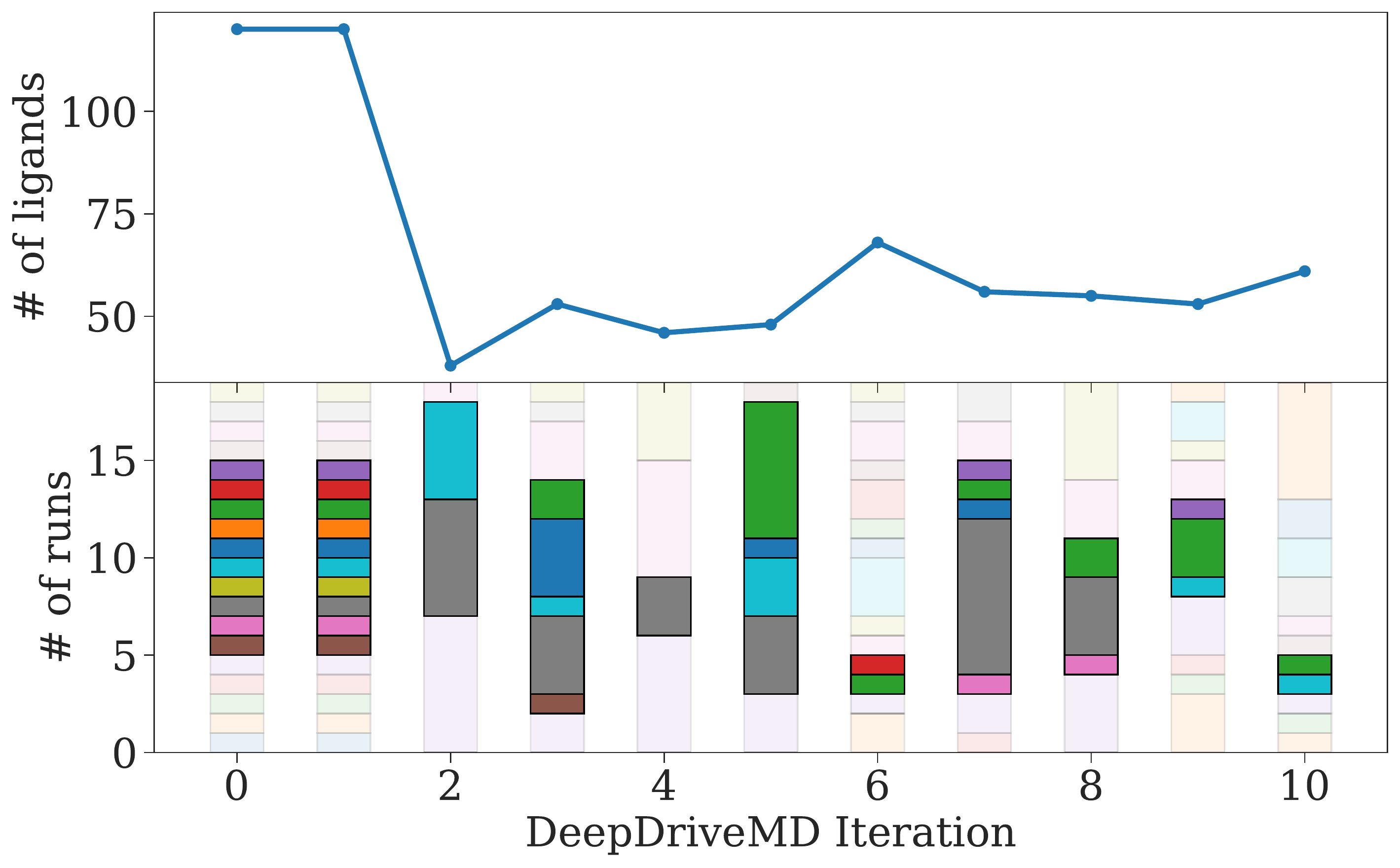}
  \caption{Top panel: DeepDriveMD-S successfully filters the initial set of 120 ligands to approximately 60 based on the uniqueness of the conformational states sampled from each of the independent simulations. The outlier selection strategy prefers ligands that induce stable interactions with PLPro, which is learned by the CVAE. DeepDriveMD-S focuses on the most impactful ligands, and brings back more candidates once the current samples are sufficiently examined. Bottom panel: DeepDriveMD-S iterations highlighting the selections of subsequent sets of ligands for a small subset of ligands (ids 5--15) of the 120 original ligands). Each bar is colored uniquely according to the ligand id. Note how the states selected constitute only ligands 7 and 9 in iteration 2, whereas at the end of iteration 10, ligands 9 and 12 are selected as part of the sampling, thus pruning out unproductive trajectories from previous iterations.}
  \label{fig:uc2_ligand_per_itr}
\end{figure}

In contrast to UC1, for which we have access to long time-scale simulations of protein folding, there are currently no baseline simulations against which we can compare the PLC simulations of UC2. Hence, we provide here only a qualitative evaluation of UC2, in which we assess the effectiveness of conformational sampling by examining how DeepDriveMD-S `prunes' unproductive simulations based on the criterion of protein-ligand interaction stability. As each iteration of DeepDriveMD-S proceeds, one would expect that productive trajectories would improve the stability in binding---thus, pruning away PLCs that exhibit weak (or no) interactions. 

We utilize a similar model, namely the CVAE that learns to represent salient features implicated in PLPro-ligand binding. We initially started with 120 ligands (extracted from a virtual screening study), whose effects on PLPro are initially unclear. We expect that after the first round of learning, inference, and simulation restarts, DeepDriveMD-S will be able to filter out ligands that do not potentially interact with PLPro. The outlier detection method, in this case a combination of DBSCAN and LOF, works to pick out and rank the data points that are farthest from the mean interaction profile for the PLCs. These points in latent space represent the novel states that are less sampled by MD simulations. From them, we spawn new simulations to enhance sampling around these areas of conformational space, and at the same time identify the ligands that can improve the stability of interactions with  PLPro. 

The top panel in \autoref{fig:uc2_ligand_per_itr} provides an illustration of how the initial 120 ligands are pruned to the top 40 within the first two iterations of DeepDriveMD-S. In successive iterations, DeepDriveMD-S automatically  reevaluates and extends the search space by suggesting restart points from the previously ignored ligands, increasing the number of simulated ligands back to approximately 60. To provide an illustration of this pruning process, we consider a subset of about 10 ligands in the bottom panel of \autoref{fig:uc2_ligand_per_itr}. For example, amongst ligands 5--15, 8 out of 10 ligands are sampled after iteration 1. The workflow then hones in on ligands 7 and 9 at iteration 2, and then returns to consider ligands 5 and 12 in the later iterations. However, at the end of iteration 10, only two ligands remain, indicating that these constitute the most interesting PLCs within this subset. While we note that this provides a qualitative evaluation of DeepDriveMD-S in UC2, our results reinforce  that using ML/AI-driven techniques can filter potentially unproductive trajectories, maintaining a balance between exploration vs. exploitation. In \autoref{sec:performance}, we present results in characterizing the performance of  DeepDriveMD-F, where a slightly modified version of the CVAE is used, namely the three-dimensional adversarial autoencoder (3dAAE)~\cite{casalino2020aidriven}. This model allows us to overcome implementation challenges (since the CVAE's space and time complexity is quadratic in the number of amino-acid residues, whereas the 3dAAE is linear) while being able to test the performance of the ML/AI approach on the ORNL Summit system.   

\subsection{Scientific Performance}
\begin{figure}[!htpb]
  \centering
  \includegraphics[width=\columnwidth,trim=2.5mm 3mm 3mm 4mm,clip]{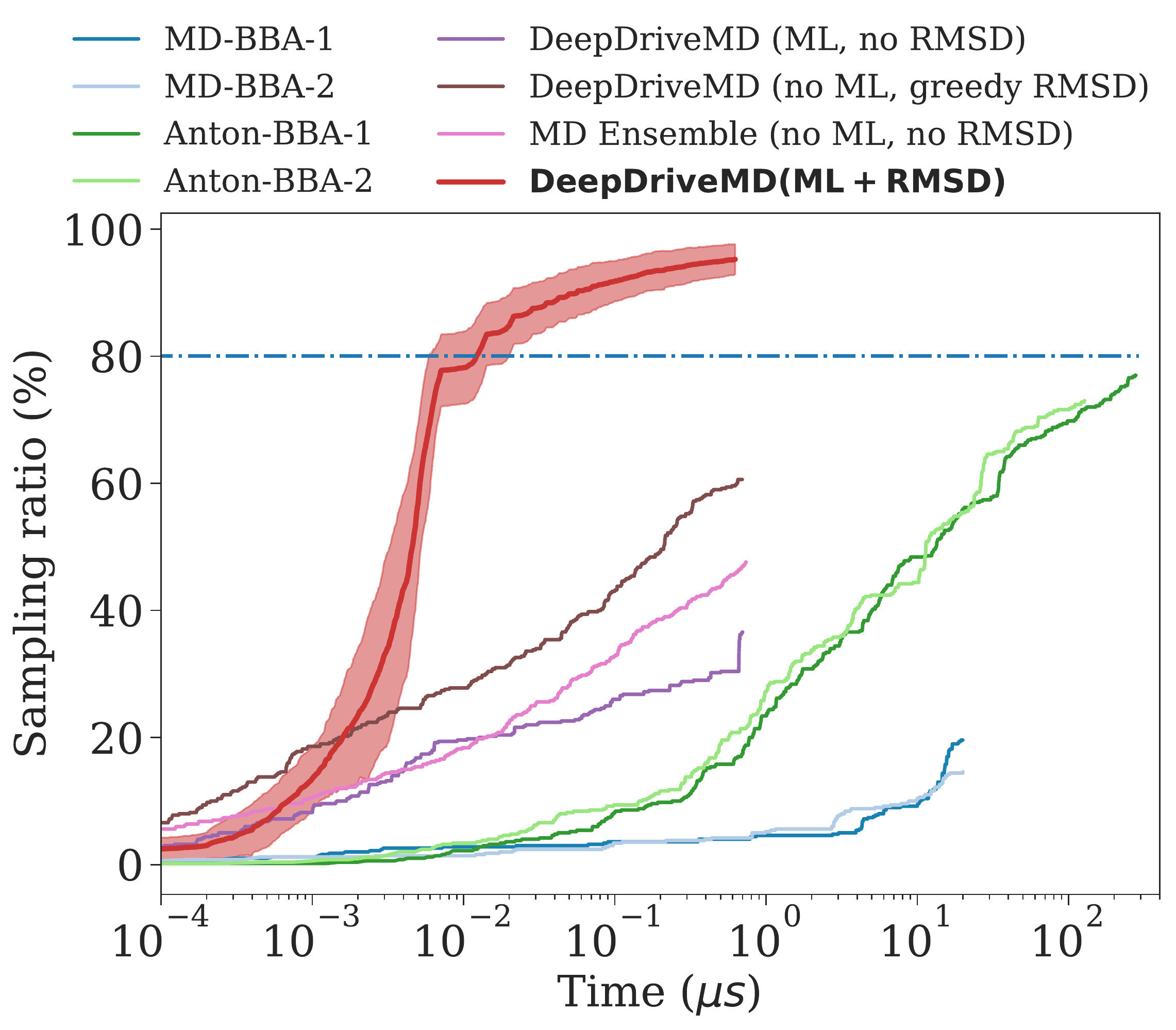}
  \caption{Sampling ratio of BBA conformational states as a function of
    simulation time.  DeepDriveMD samples conformational states more effectively
    than free MD simulations (including those run on Anton-1). When RMSD is
    used to filter the ML-selected outliers, sampling performance is better than
    when using either RMSD or ML alone. Compared to the Anton simulations, the
    ML + RMSD strategy (for which uncertainty from 10 trials is shown in light
    red) reaches 80\% sampling more than 1000$\times$ faster.}  
  \label{fig:pf-states}
\end{figure}

We measure the \emph{sampling effectiveness} as the fraction of the total population explored---with and without ML approaches---as a function of time~\cite{hruska2019extensible}. To achieve this, we selected the states sampled at some time and compare it against all states obtained with respect to reference simulations. For UC1, we compare our simulations to $O$(100~$\mu$s) simulations of the BBA system performed with the Anton-1 hardware, providing extensive information on all possible states sampled for its folding process~\cite{Lindorff-Larsen517}. The conformers from the these simulations are embedded into a 10-dimensional CVAE latent space. These embeddings are then clustered with MiniBatchKmeans, with $k$=500. This setup allows us to compare the states sampled by each simulation, while keeping a consistent definition of the conformational states sampled. In \autoref{fig:pf-states}, the ratio of sampled states in a simulation at a time $T$ is defined as the number of clusters that the simulation has traversed by that time, divided by the total number of clusters (500)~\cite{hruska2019extensible}. A potential limitation of this approach is the constrained total number of states considered. Usually, the conformational states in a simulation are aggregated (typically by using clustering approaches to merge smaller/ less-populated states) to describe a compact representation of the sampled landscape~\cite{Mardt_2018}. In the work reported here, however, we discretized the states to a number of clusters that reflects a tradeoff between aggregating too many states (that are much smaller in population) versus breaking up larger ones. 

\autoref{fig:pf-states} indicates that our ML-driven methods enhance sampling by a factor of more than 1000: It takes the ML-driven ensemble around 10~ns of aggregate simulated time (over 120 simulations) to cover 80\% of the 500 conformational states, while the Anton simulations take at least 100~$\mu$s to cover a somewhat lower fraction. To provide a comparison with similar hardware (i.e., similar GPUs used for simulations and same simulation conditions), we observe that a single MD simulation executed on the same GPU for 12 hours (blue lines; MD-BBA-1 and MD-BBA-2) sample less than 20\% of the states; whereas an MD ensemble (no ML, no RMSD based selection; light pink line) run on 120 GPUs samples close to 50\% of the states. 

\begin{figure}[!htpb]
  \centering
  \includegraphics[width=\columnwidth,trim=3mm 3mm 3mm 4mm,clip]{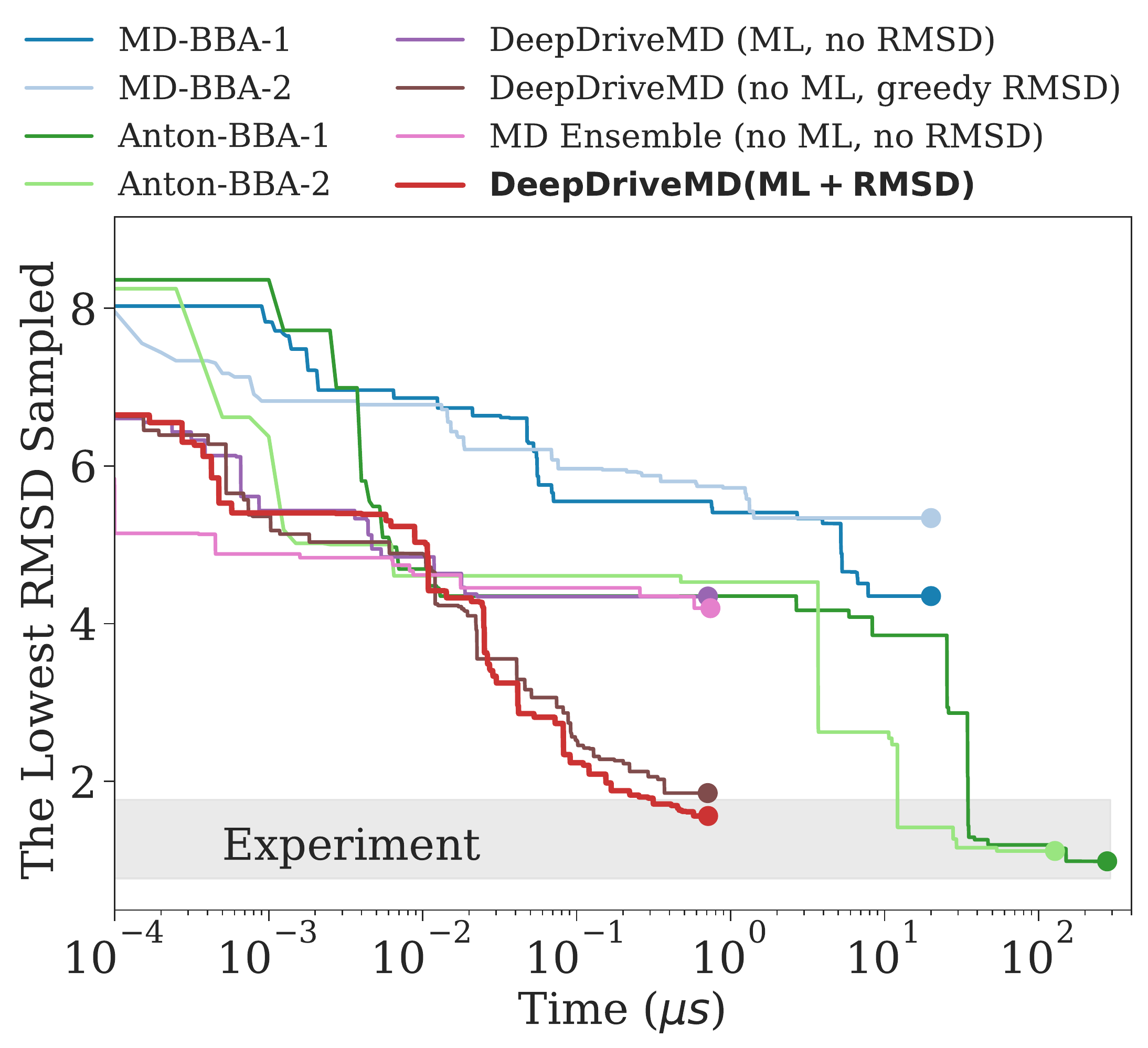}
  \caption{Lowest RMSD to BBA's folded state sampled, as a function of simulation time. 
    The gray shaded area  depicts the intrinsic conformational diversity in the NMR ensemble (average RMSD 1.3 $\pm$ 0.5 \AA).
  Each of the three DeepDriveMD variants, and the MD ensemble, runs for 12 hours on 120 Lassen nodes. DeepDriveMD accelerates sampling along the RMSD progress coordinate, compared to free MD simulations (including those run on Anton). The No ML, greedy RMSD strategy quickly filters out the high RMSD states, but its subsequent sampling is stagnant. The ML + RMSD approach proceeds more consistently and eventually overtakes the greedy strategy.}
  \label{fig:rmsd_bba}
\end{figure}

\autoref{fig:rmsd_bba} provides further evidence that DeepDriveMD's ML-driven strategy accesses folded state conformations similar in quality to those identified by the Anton-1 simulations, despite running for only 12 hours. The DeepDriveMD simulations access the shaded region, representing intrinsic conformational diversity as revealed by experiment, when using the ML + RMSD strategy, and comes very close when using the no ML, greedy RMSD strategy. The latter quickly filters out the high-RMSD conformations to reach low-RMSD states, but without the access to underlying conformational information about sampled states, cannot filter out the local minima in the energy landscape, which curbs its sampling efficiency due to trapping in some intermediate states; thus, it is eventually overtaken by the ML + RMSD strategy. This result supports the idea that ML + RMSD-based sampling provides a significant boost to the overall sampling process. 

\section{Performance Characterization}
\label{sec:performance}
We use our three use cases to evaluate various aspects of DeepDriveMD computational performance.

\subsection{Sequential vs. Concurrent Execution of Components}

The DeepDriveMD framework makes it straightforward to switch from file-based, sequential composition of stages (good for debugging) to stream-based, concurrent composition (typically more efficient). To explore the impact of these two implementation approaches, we compare the performance of two alternative implementations, one sequential (DeepDriveMD-F) and one concurrent (DeepDriveMD-S). We show in Table~\ref{tab:performance} the parameters used for UC1 experiments that measure the scaling behavior of the two variants on LLNL Lassen, when folding the BBA protein on 30 and 33 compute nodes, with DeepDriveMD-F and DeepDriveMD-S respectively. Measuring overheads and resource utilization for both experiments enables a direct comparison between the performance of the file- and streaming-based coordination patterns for UC1. We see clear indication that DeepDriveMD-S outperforms DeepDriveMD-F, producing 1.5X more MD data per hour and running 8.4X and 74.9X more training and inference iterations per hour, respectively. The concurrency of DeepDriveMD-S therefore allows the model to learn more (50 training epochs vs 20) from generated data, and make online decisions to steer sampling with a larger and more current pool of outliers (1000-5000 vs 500-700) than DeepDriveMD-F, closing the continual learning loop as fast as possible. 

As DeepDriveMD-S uses streaming communication, it enables increased concurrency among tasks and avoids I/O bottlenecks due to concurrent tasks writing files. In contrast, within DeepDriveMD-F, each simulation task writes two files, one containing the simulation trajectory and one for preprocessed data, creating pressure on the network file system which increases with the number of concurrent simulation tasks, thus not being able to accommodate larger simulation systems as a consequence of the I/O overheads.

\begin{figure*}
  \centering
  \hrule height 0.3mm
  \includegraphics[width=\textwidth,trim=0.5mm 2.2mm 0.4mm 1.8mm,clip]{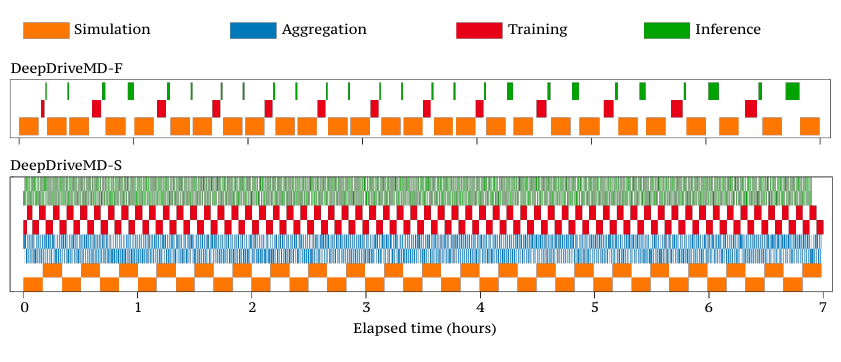}
  \hrule height 0.3mm
  \vspace{1.5ex}
  \caption{Task execution timeline when running UC1 on Lassen. Each bar corresponds to one iteration of a Simulation, Aggregation, Training, or Inference task. DeepDriveMD-F runs training after every other iteration while DeepDriveMD-S runs each type of task concurrently and without gaps. In DeepDriveMD-S, every second iteration is shown on a different line to distinguish between different iterations. For clarity, only 1 of 120 Simulation tasks and 1 of 10 Aggregation tasks are shown for DeepDriveMD-S.}
  \label{fig:streaming_gantt}
\end{figure*}

In the Simulation stage of both implementations, 120 MD simulation tasks run in parallel. However, as one can see from \autoref{fig:streaming_gantt}, in DeepDriveMD-S simulations run continuously without gaps, while in DeepDriveMD-F the Simulation stage is sequentially followed by a Training stage (every other iteration) and an Inference stage. As a result, DeepDriveMD-S performs 1.5 times more simulations per unit of time than does DeepDriveMD-F (orange bars in \autoref{fig:streaming_gantt}). 

While DeepDriveMD-F avoids executing Aggregation tasks given a feature which makes this optional, \autoref{fig:streaming_gantt} clearly shows that DeepDriveMD-S executes many more Simulation, Training, and Inference tasks in the seven-hour period than does DeepDriveMD-F. In addition to enabling simulation tasks to run continuously, DeepDriveMD-S permits the analysis of partial simulation data to proceed while simulations are executing, further speeding up the execution of the overall workflow compared to DeepDriveMD-F.

\begin{table*}[!bht]
  \caption{DeepDriveMD-F (“F”) and DeepDriveMD-S (“S”) performance and resource utilization for UC1 on Lassen. The application executes on 30 (for F) or 33 (for S) compute nodes. Both run 120 10~ns Simulation tasks on 30 nodes, each using 1 CPU and 1 GPU. S runs 10 Aggregators on one node, with each using 1 CPU; F has no Aggregator. Both run one Training task that uses 1 CPU and 1 GPU, on a separate node for S, and on a Simulation node for F. Both run one Inference task that uses 39 CPUs and 1 GPU on a separate node for S, and 1 CPU and 1 GPU on a Simulation node for F. ``It"=iteration.}
  \label{tab:performance}
  \centering
  \small
  \begin{tabular}{c | cc | ccc | cccc | cccc}
    \toprule
    \multirow{2}{*}{\textbf{System}}                   &
    \multicolumn{2}{c|}{\textbf{Simulation}} &
    \multicolumn{3}{c|}{\textbf{Aggregation}}   &
    \multicolumn{4}{c|}{\textbf{Training}}   &
     \multicolumn{4}{c}{\textbf{Inference}}   
             \\
                         &
    \textbf{time}    &
    \textbf{iter/h}   &
    \textbf{time}    &
    \textbf{it/h}   &
    \textbf{tasks}  &
    \textbf{time}        &
    \textbf{it/h}       &
    \textbf{sample/it} &
    \textbf{epochs}        &
    \textbf{time}        &
    \textbf{it/h}       &
    \textbf{sample/it} &
    \textbf{outliers}      \\
    \midrule
    F  &
    591 s          &
    3.9            &
    N/A            &
    N/A            &
    N/A        &   
    282 s          &
    2.0              &
    24K; 48K       &
    15; 20         &
    111 s          &
    3.7            &
    24K; 48K &
    500--700       \\

    S  &
    576 s          &
    5.9            &
    3.2 s          &
    1091       &
    10             &
    216 s          &
    16.7            &
    20K            &
    50             &
    13 s          &
    277             &
    20K            &
    1000--5000     \\
    \bottomrule
  \end{tabular}
\end{table*}

DeepDriveMD-S therefore uses computing resources more effectively than DeepDriveMD-F, which periodically leaves some nodes idle. For example, while DeepDriveMD-F runs 120 simulation tasks concurrently on 120 GPUs during the Simulation phase, it only uses one GPU for the training task, thus leaving 119 GPUs idle during that stage. Furthermore,  DeepDriveMD-S can avoid saving intermediate data to disk by using the network via ADIOS SST to communicate among its components. For UC1, DeepDriveMD-S avoids writing $\sim$50GB to disk by streaming data among the concurrently executing tasks. This is especially important for larger and more complex biological systems~\cite{casalino2020aidriven,Dommer2021}, such as UC3, where data from large simulations across $\sim$100 nodes would create an I/O bottleneck without this streaming capability.

We estimate the overall percentage of time spent on ADIOS I/O to be 0.8\%, by adding ADIOS I/O times from all tasks and dividing it by the total wall time over all the tasks. Moreover, most of this time is hidden from the end user due to the fact that the components run concurrently. While the ADIOS overhead in simulations is 0.3\%, the benefit of continually running the simulations greatly outweighs the communication cost. In addition, the performance of the other components might affect the sampling efficiency but not the total number of executed simulations, due to the loose coupling of tasks.

\subsection{DeepDriveMD-F performance}

Table~\ref{tab:exp} shows the parameters used for the UC2 scaling study. These experiments show the overheads and resource utilization of DeepDriveMD-F when varying four configuration parameters: (1) the HPC platform; (2) the number of ligands; (3) the number of tasks executed; and (4) the amount of resources requested. Experiments PLC-1--4 evaluate between one and eight ligands, using between 120 and 960 GPU devices on Summit and Lassen. Experiment PLC-5 evaluates 51 ligands, using 6120 GPUs (1020 nodes) on Summit, while experiments PLC-1e and 4e use 120 and 960 GPUs (20 and 160 nodes) to evaluate 120 and 960 ligands, respectively.

\begin{table}
  \caption{Configuration \& overhead for UC2 experiments. Overheads are always low relative to a typical 12-hour runtime.}
  \label{tab:exp}
  \centering
  \small
  \begin{tabular}{lrrrccl}
    \toprule
    \textbf{Exp.}            &
    \textbf{Ligands}          &
    \textbf{GPUs}            &
    \textbf{Tasks}            &
    \textbf{Platform}           &
    \textbf{Overhead} \\
    \midrule
    PLC-1                 &  
    1                     &  
    120                    &  
    250                   &  
    Summit                &  
    334.2~s               \\ 
    PLC-2                 &  
    1                     &  
    120                    &  
    250                   &  
    Lassen                &  
    302.3~s               \\ 
    PLC-3                 &  
    8                     &  
    960                   &  
    2000                  &  
    Summit                &  
    265.1~s               \\ 
    PLC-4                 &  
    8                     &  
    960                   &  
    960                   &  
    Lassen                &  
    314.8~s               \\ 
    PLC-5                 &  
    51                    &  
    6120                  &  
    6120                  &  
    Summit                &  
    254.0~s               \\ 
    \midrule
    PLC-1e                 &  
    120                   &  
    120                    &  
    120                   &  
    Summit                &  
    325.8~s               \\ 
    PLC-4e                &  
    960                   &  
    960                   &  
    960                   &  
    Summit                &  
    376.0~s               \\ 
    \bottomrule
  \end{tabular}
\end{table}

In the PLC experiments, we count as overhead all time spent not executing any workflow task when resources are available; this includes the time taken by DeepDriveMD-F and RADICAL-Cybertools to resolve task dependencies, prepare the execution environment, and submit tasks for execution. We observe that overhead is relatively independent of the number of ligands analyzed, averaging $\sim$310 seconds across the PLC-x experiments. As the PLC-x experiments have different total execution times, we conclude that overheads are also invariant of the time taken by all workflow tasks to execute. We also observe that overheads are comparable between Summit and Lassen, suggesting that they are not platform-dependent. These observations suggest that DeepDriveMD-F scaling behavior is decoupled from the configuration of the use cases it supports, the time taken to execute the workflows, and the HPC platform on which it executes. The $\sim$310 seconds of overhead are insignificant relative to the 12 hours runtime of a typical DeepDriveMD UC2 run. Furthermore, these overheads all but vanish for DeepDriveMD-S, since once its tasks are launched they execute continuously through the duration of the workflow.

\subsection{DeepDriveMD-S I/O performance}

The BBA and PLPro biomolecular systems analyzed in UC1 and UC2 are relatively small: just 28 and 309 residues, respectively. The much larger 3375 residue SARS-CoV-2 spike protein system studied in UC3 results in the communication time between the Aggregation component and the Training and Inference components becoming prohibitively large, in some cases even exceeding computation time. Thus we made two small modifications to DeepDriveMD-S to optimize I/O for large biomolecular systems. First, we converted the ADIOS BP file to network communication between components. This change was simple, requiring just a change to the ADIOS XML configuration file and minimal modifications to the code (recall that simulations and aggregators were already communicating via ADIOS network streams). Second, we turned off compression of contact maps. Because the contact maps produced by the simulations are large, by default we compress them in the Simulation component and, after transfer, decompress them in the Training and Inference components. However, while we are able to achieve a lossless compression factor of 16, the costs of compression and subsequent decompression were (at least in our implementation) inordinately high in the 3375 residue case. 

Using network communication and avoiding compression delivered dramatic improvements in I/O performance, as summarized in Table~\ref{tab:UC3io}. Comparison of time reductions achieved by each of the two modifications in turn show that it is the second, eliminating compression, that delivers the largest time reduction. We leave further optimization and analysis of UC3 to future work. 

\begin{table}
  \caption{Average read times per iteration in Training and Inference components (seconds) for UC3, before and after optimizing contact map compression and communication methods.}
  \small
  \label{tab:UC3io}
  \centering
  \begin{tabular}{lrr}
    \hline
    & Before & After \\ \hline
    Training & 1464 $\pm$ 78 s & 9 $\pm$ 4 s\\
    Inference & 2239 $\pm$ 20 s & 12 $\pm$ 2 s\\ \hline
  \end{tabular}
\end{table}

\section{Conclusions}
\label{sec:conclusion}
We have described DeepDriveMD, a general-purpose and extensible framework for implementing ML/AI-driven simulation applications. We used three biophysical MD modeling applications to evaluate its design, implementation, and performance, and demonstrate that by driving ensembles of MD simulations with ML approaches, DeepDriveMD can achieve between 10--1000x improvement in time-to-solution relative to non-ML-driven approaches. For a protein folding simulation, DeepDriveMD achieves 1000$\times$ acceleration, while covering the same conformational landscape as quantified by the states sampled during a simulation. For UC2 and UC3, we  presented insights into the performance trade-offs involved in managing such diverse workloads. 

DeepDriveMD uses RADICAL-Cybertools abstractions and mechanisms to manage challenging workloads, involving diverse mixes of simulation and AI phases, on some of the largest HPC platforms available. Its support for concurrent execution of different phases, with streaming of data between components implemented with ADIOS, permits high parallel efficiency and performance.
By insulating both scientific practitioners and scientific algorithms and methods developers
from important but otherwise irrelevant details, it allows domain scientists and 
methods developers alike to advance scientific discovery on
high-performance platforms.

ML methods play an increasingly visible and important role in computational modeling due to their ability to enable smarter computational campaigns and thus accelerate scientific discovery. Such methods are successful because they offer simple, scalable, and fairly general means to deal with high-dimensional, potentially high volume and velocity scientific datasets---a capability that is particularly important when working with biomolecular systems due to the high dimensionality of the simulation datasets. As recent applications also make clear~\cite{Dommer2021}, the ability to interface emerging AI-hardware with ensemble MD simulations promises even greater benefits for accessing time- and length-scales longer than conventionally possible. DeepDriveMD thus unifies and permits the large-scale use of ML-driven simulation methods for such applications.

\vspace{1ex}

\noindent
\small{\emph{Acknowledgements}:}
\small{This article reports on work supported by the Exascale Computing Project (17-SC-20-SC), a
collaborative effort of the U.S.\ Department of Energy (DOE)'s Office of Science and National Nuclear
Security Administration:
in particular, ECP projects CANDLE, CODAR, and ExaWorks. This research used resources at the Oak Ridge Leadership Computing Facility at the Oak Ridge National Laboratory, which is supported by the DOE Office of Science  under Contract No. DE-AC05-00OR22725; the Argonne Leadership Computing Facility (ALCF), a DOE Office of Science User Facility supported under Contract DE-AC02-06CH11357; and Livermore Computing at the Lawrence Livermore National Laboratory. Funding for this work was provided by the Department of Energy’s Advanced Scientific Computing Research Program through grant number 31975.2 to Argonne National Laboratory for the Co-Design of Advanced Artificial Intelligence (AI) Systems for Predicting Behavior of Complex Systems Using Multimodal Datasets project.
Anda Trifan acknowledges support from the DOE through the Computational Sciences Graduate Fellowship (DOE CSGF) under grant number DE-SC0019323.}

\small
\bibliographystyle{IEEEtran}
\balance
\bibliography{Bibs/main,Bibs/radical,Bibs/DDMD-refs,Bibs/refs-NAMD,Bibs/biophys-refs}

\end{document}
\endinput